\documentclass[]{emulateapj}

\newcommand{\Msun}{${\rm M_\odot}$}
\newcommand{\Rsun}{${\rm R_\odot}$}

\newcommand{\RpRs}{$R{\rm _p}/R{\rm _*}$}

\slugcomment{ApJ, in press}

\received{2009 Feb 23}
\accepted{2009 Mar 12}
\begin{document}

\title{Photometric Detection of a Transit of HD 80606b}

\author{
E.~Garcia-Melendo\altaffilmark{1} \&
P.~R.~McCullough\altaffilmark{2}
}

\email{egarcia@foed.org}

\altaffiltext{1}{Esteve Duran Observatory Foundation, 08553 Seva, Spain}
\altaffiltext{2}{Space Telescope Science Institute, 3700 San Martin Dr., Baltimore MD 21218, USA}

\begin{abstract}
We report a times series of B-band photometric observations initiated on the eve
of Valentine's day, February 14, 2009, at the
anticipated time of a transit of the extrasolar planet HD 80606b. 
The same transit has been observed independently (Fossey et al. 2009; Moutou et al. 2009).
For one transit model favored by the data, minimum light 
equals 0.990 times the nominal brightness of HD 80606 and occurs at HJD 2454876.33.
The latter time, combined with the orbital period $P = 111.4277 \pm 0.0032$ days,
longitude of periastron, $\omega = 300.4977 \pm 0.0045$ degrees,
and time of mid-secondary eclipse HJD $2454424.736 \pm 0.003$ (Laughlin et al. 2009), 
refines the orbital eccentricity and inclination.
The duration of the model transit is 0.47 days,
and its four contacts occur at HJD 2454876 plus 0.10, 0.24, 0.42, and 0.57 days.
We describe parameterizations of a transit model with mutually accommodating eccentricity,
$e = 0.9337^{+0.0009}_{-0.0006}$, inclination, $i = 89.26^{+0.24}_{-0.09}$ degrees,
and the planetary radius in units of the stellar radius \RpRs$ = 0.11^{+0.04}_{-0.02}$.
\end{abstract}


\section{Introduction}

HD 80606b is a gaseous giant planet, four times the mass of Jupiter, in an eccentric
($e = 0.93$) 111-day orbit about the G5V star HD 80606 (Naef et al. 2001). 
HD 80607, a physical companion to HD 80606 (Naef et al. 2001), provides a convenient
comparison star for differential photometry of HD 80606, because the two have
similar brightnesses, colors, and are separated by only $\sim$20\arcsec. 
Laughlin et al. (2009) demonstrated with {\it Spitzer} IRAC photometry at 8-$\mu$m that
the planet HD 80606b passes behind its host star and estimated a 15\% probability
that HD 80606b passes in front of its star also.
With allowance for $3\sigma$ uncertainty in the eccentricity estimated from radial velocities,
$e = 0.9327 \pm 0.0023$ (Laughlin et al. 2009), 
and for a maximum half-duration of $\sim 0.36$ day for a central transit, 
the window of opportunity to observe either ingress or egress was HJD $2454876.44 \pm 1.05$ day. 
At www.oklo.org, Laughlin encouraged a global campaign of observations 
near the anticipated time of transit of HD 80606b.

\section{Observations}

We observed the pair of stars HD 80606/7 with the 0.6-m diameter
Cassegrain telescope of the Esteve Duran Observatory.
The 16-bit camera is a ST-9XE model manufactured by the Santa Barbara Imaging Group.
For all of the observations reported here, we used an Optec B band filter.
The CCD has 512 pixels by 512 pixels; each pixel is 20~$\mu$m~$\times$~20~$\mu$m,
or 1{\arcsec}.27~$\times$~1{\arcsec}.27.
Images were corrected with dark images and flat fields.
Individual exposures were 10 seconds, obtained at a cadence of 12.1 seconds.
The B filter enabled keeping the shutter open for a large fraction (83\%) of the
observing time while avoiding saturation from the bright stars. 
The point spread function (PSF) of HD 80606 had a peak of $\sim$8000 e$^-$ pixel$^{-1}$
and an integral of $\sim$90000 e$^-$; the background was $\sim$100 e$^-$ pixel$^{-1}$.
The PSF's FWHM was $\sim$5\arcsec, which
prevented blending of the wings of the PSFs of the
target, HD 80606, and the comparison star, HD 80607.

\section{Analysis}

\begin{figure}
\plotone{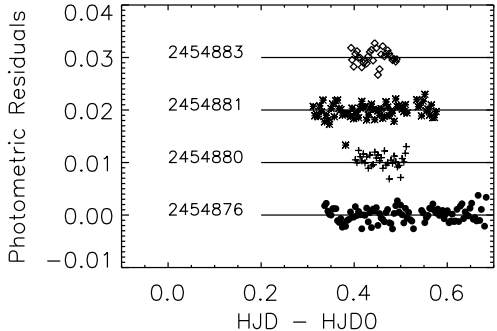}
\caption{Photometric residuals. 
Differential photometric residuals of HD 80606 with respect to HD 80607 on four nights
are shown. 
The four times series were obtained and analyzed in an identical manner.
The original observations have been averaged into 5-min bins.
The residuals from the night of the transit (HJD0 = 2454876) 
are with respect to the
nominal transit model A (Figure \protect{\ref{fig:bestlc}}).
The others, the ``controls,'' 
are with respect to constant-brightness models on the dates indicated and have been
shifted vertically for clarity.
The values of HJD0 are indicated to the left of each series.
On this scale, the observed transit would extend vertically from one horizontal line
to another, i.e. by 1\% in flux.
\label{fig:controls}}
\end{figure}

Because the angular separation of the two stars
was $\sim$4 times their PSF's FHWM, we used conventional aperture photometry.
Results do not depend significantly on aperture size; the results reported
here used a circular aperture of radius 7 pixels ($\sim9${\arcsec}).
The small angular separation, similar color
indices between the target star and the comparison star,
and moderate range of airmass (1.0 to 1.5) permitted differential photometry
with very small corrections for differential color and differential extinction. 
Using nominal wavelength-dependent atmospheric extinction 
(Table 5 of Hayes \& Latham 1975), blackbody approximations
to the stellar spectra of HD 80606 and HD 80607, and their
effective temperatures, 5645 K and 5555 K respectively
(Naef et al. 2001), we corrected the B-band, differential photometry of
HD 80606 with respect to HD 80607 by subtracting 0.5 mmag per airmass.
At a nominal B-band extinction of 0.31 mag airmass$^{-1}$, and with the maximum
differential airmass between the two stars equal to $1.6\times 10^{-4}$,
differential extinction is not more than 0.05 mmag, so we neglect it.
The nightly photometric averages
of out-of-transit time series differ by $\sim$1 mmag for
four different nights and are consistent with a
very small correction for differential color and a negligible correction
for differential extinction.
The night of the transit, the sky
was clear all night; on three subsequent nights used to obtain ``controls,''
the clarity was substantially more variable, and yet the differential
photometry remained similar in quality to that of the night of the transit,
0.63\% r.m.s. per 10-sec exposure, and 0.14\% r.m.s. per 5-min average (Figure \ref{fig:controls}).

The stars drifted $\sim$60 pixels across the CCD during the first half of the night
of the transit; hereafter we refer to that drift as the ``nominal'' trajectory. 
The drift has no measurable effect on the differential photometry.
After a 17-min gap in observations required during
meridian crossing (from HJD 2454876.531 to 2454876.543), the stars were re-acquired
again near their positions at which
they began the night. Although the positions have a discontinuity across the gap of
50 pixels, the differential photometry does not change across the gap.
Indeed, zero offset has been applied to match the photometry across the gaps at
the meridian in any of the light curves. 
Serendipitously the vector offset from HD 80606 to HD 80607 is similar in direction
but $\sim$1/4 as large as the 60-pixel length of the nominal trajectory; consequently,
any gradient of erroneous calibration across the {\it entire} trajectory
(e.g. due to a gradient in either vignetting or quantum efficiency of the CCD),
would be reduced by $\sim$4x in the {\it differential} photometry.
The B-band flat field is very uniform; along the nominal trajectory, there is
a single feature, a ring from a de-focused dust speck, with a 4-pixel inner radius
and an 8-pixel outer radius, and $\la$0.8\% fainter than nominal within the
two radii. 
The flat field spatial variation is 0.6\% r.m.s. on a per-pixel basis, 
with a best-fitting gradient that amounts to 0.4\% from beginning-to-end across the
nominal trajectory. Even if the latter gradient were erroneous, and we have no reason to believe
it is, then it would only amount to $\sim$0.1\% error in photometric calibration, or $\sim$10\%
of the depth of the transit.
On the final ``control'' night, we purposefully
induced a drift of the stars' positions on the CCD
along a trajectory similar in initial position and vector direction to the nominal
trajectory.
Even with an induced drift of more than twice the angle of the nominal
trajectory, we detected no significant correlation of differential photometry
with image position on the final ``control'' night.

\begin{figure}
\plotone{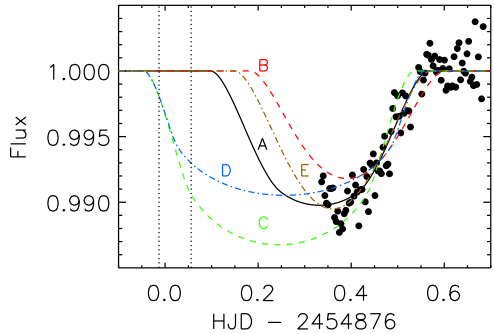}
\caption{Egress of HD 80606b.
Time-series B-band differential photometry of HD 80606,
as observed with the 0.6-m telescope of the Esteve Duran Observatory 
during the February 13-14, 2009 transit of HD 80606b.
The data from Table 1, averaged into 5-minute intervals (filled circles),
and five B-band models are shown:
the nominal transit model A with \RpRs\ = 0.11 (solid line),
models B and C also with \RpRs\ = 0.11 (dashed lines),
and
models D and E with \RpRs =0.09 and 0.15, respectively (dot-dashed lines).
Each model has an associated eccentricity and inclination (Table 2 and Figure \ref{fig:chisq}).
Models B, C, and D are all marginally-acceptable in matching the B-band observations.
Models C and D are only marginally consistent with the 
non-detections at HJD~$<$~2454876.00 of the transit ingress.
Model E matches the B-band observations as well as model A but
model E has a marginally-acceptable minimum light at R-band (not plotted; see text
and Table 2).
Terminations of time series obtained with the DEMONEX telescope (Eastman et al. 2009) and one of the MEarth telescopes (Irwin et al. 2009)
are indicated by dotted lines at 2454875.987 and 2454876.056 respectively.
\label{fig:bestlc}}
\end{figure}

We model the observed transit light curve (Figure \ref{fig:bestlc})
using the algorithms of Mandel \& Agol (2002). A required input to the latter
algorithms is the time-dependent projected
separation of the center of the planet from the center of the star in units of the
radius of the star. For the latter, we use Eq. 5.63 of Hilditch (2001) for eccentric
orbits.
For simplicity, we interpolate and fix the quadratic limb darkening coefficients,
$\gamma_1 = 0.711$ and $\gamma_2 = 0.124$, 
from Claret (2000) as appropriate for the B filter bandpass, 
and for HD 80606's spectroscopic gravity, effective temperature, and metallicity
(logg = $4.50\pm0.20$, $T_{\rm eff} = 5645\pm45$, and [Fe/H] = $+0.43\pm0.06$; Naef et al. 2001).
We adopt the following orbital parameters from Laughlin et al. (2009):
the orbital period $P = 111.4277 \pm 0.0032$ days;
the longitude of periastron, $\omega = 300.4977 \pm 0.0045$ degrees,
and the heliocentric Julian date of periastron passage, $T_{\rm peri} = 2454424.86$.
We adopt the dimensionless semi-major axis $a/R_* = 100$,
which is equivalent to $M_* = 0.98~$\Msun\ and $R_* = 0.98~$\Rsun.

\begin{figure}
\plotone{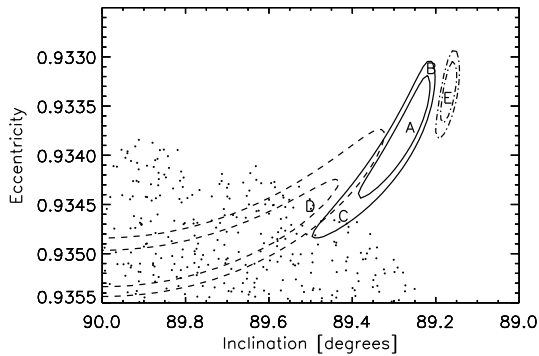}
\caption{Contours of reduced chi-squared for the transit light curve (Figure \protect{\ref{fig:bestlc}}).
Letters correspond to models of Table 2.
The letter A indicates the nominal model adopted in this work.
As indicated by the contours and discussed in the text, 
for a specific value of \RpRs, deviations from the nominal
value of inclination can be compensated with deviations in eccentricity.
Larger values of \RpRs\ require smaller inclinations.
The dashed, solid, and dot-dashed contours correspond to \RpRs\ =
0.09, 0.11, and 0.15, respectively. In each case, the inner and outer contours
correspond to $\chi^2_\nu =$ 1.5 and 2.0, respectively.
The orientation of the axes makes the pattern of the locations of the letters match that
of Figure \protect{\ref{fig:bestlc}}. The contours are limited primarily by the
observed time of egress; their arc-like shapes are caused by the star's circular edge.
The stippled region is excluded by non-detections of the transit ingress; its
periphery is an inverted arc caused by the opposite edge of the star.
\label{fig:chisq}}
\end{figure}

There are three remaining parameters required in order to
model the light curve: the ratio of planetary and stellar radii (\RpRs),
the eccentricity,
and the orbital inclination. Because we only observed egress, these three
parameters are correlated (Figure \ref{fig:chisq}), and their uncertainties
are asymmetric. 
We cannot distinguish between a model with a
smaller planet, larger inclination, and larger eccentricity from another model
with a larger planet, smaller inclination, and smaller eccentricity.
Figure \ref{fig:bestlc} shows five models with (\RpRs, $e$, $i$) triplets itemized in Table 2. 
The nominal model A has \RpRs$ = 0.11$, $e = 0.9337$ and $i = 89.26$\arcdeg, and
reduced chi-squared $\chi^2_\nu = 1.0$.
The associated minimum projected separation during transit of
the centers of the planet and the star, $z_{\rm min} = 0.85 R_*$.
Also with \RpRs$ = 0.11$, models B and C are at the limits of eccentricity permitted by
our data; model C also is only marginally consistent with the non-detections
at HJD~$<$~2454876.00 of the transit ingress by Eastman (2009) and Irwin (2009).
In the case of Model D, with \RpRs$ = 0.09$, our
data {\it prefer} a central-transit model, in order to create sufficient transit depth,
but the non-detections of ingress exclude a central transit.
Model E is the other extreme, with \RpRs$ = 0.15$; it requires a grazing transit
to match the B-band data but is only marginally consistent with preliminary
reports of R-band transit depths (Fossey et al. 2009; Moutou et al. 2009).

For any specific \RpRs, and for small deviations about the nominal values,
$e = 0.9337$ and $i = 89.26$\arcdeg, the eccentricity may be
increased by ${\delta}e$ in order to accommodate an increase in the orbital inclination
of $\sim 200\arcdeg{\delta}e$.
Increasing the eccentricity by 0.0001 shifts the transit earlier by 0.01 day.
With the constraint of the observed time of egress, increasing the orbital inclination
by 0.1\arcdeg\ must be accommodated by shifting the transit mid-point earlier by 0.05 day,
i.e. increasing the eccentricity by 0.0005.
The eccentricity is derived from fitting the light curve and is essentially determined
from the elapsed time from the secondary eclipse to the transit.
With the relatively large value of \RpRs$ = 0.15$,
model E, a grazing transit, fits our light curve and sets a lower limit $i > 89.17$\arcdeg.
With moderate values of \RpRs\ = 0.09 to 0.11, transit models fit our data well, are
consistent with the two non-detections of ingress,
and set upper limits: $e < 0.9346$ and $i < 89.50$\arcdeg.
Combining all of the constraints, we estimate the eccentricity
$e = 0.9337^{+0.0009}_{-0.0006}$, the inclination, $i = 89.26^{+0.24}_{-0.09}$ degrees,
and the planetary radius in units of the stellar radius (\RpRs$ = 0.11^{+0.04}_{-0.02}$).

\begin{deluxetable*}{ccccccccc}
\tablewidth{0pt}
\tabletypesize{\small}
\tablecaption{Time-Series Photometry \label{table:photometry} }
\tablehead{\colhead{HJD} & \colhead{Weight} & \colhead{$X$} &
   \colhead{$\hat m_{6,7}$} & \colhead{$m_{7}$} & \colhead{$x_6$} & \colhead{$y_6$} & \colhead{$x_7$} & \colhead{$y_7$} }
\startdata
2454876.333779 & 1 & 1.282 & 0.0136 & 13.4154 & 240 & 282 & 224 & 278 \\
2454876.333924 & 1 & 1.281 & 0.0100 & 13.4150 & 240 & 282 & 224 & 278 \\
2454876.334068 & 1 & 1.280 & 0.0009 & 13.4095 & 240 & 282 & 224 & 278 \\
2454876.334213 & 1 & 1.280 & 0.0004 & 13.4025 & 240 & 282 & 224 & 278 \\
2454876.334357 & 1 & 1.279 & 0.0048 & 13.4009 & 242 & 282 & 226 & 278 \\
\enddata
\tablenotetext{}{Column 1 is the mid-exposure heliocentric Julian day. 
Column 2 is a weight: 1 nominal; 0 for discrepant points. 
Column 3 is the airmass $X$. 
Column 4 is the B-band magnitude of HD 80606 ($m_6$) 
with respect to the comparison star HD 80607 ($m_7$), with a differential
color correction applied: 
$m_{6,7} = m_6 - m_7 - 0.5 \times 10^{-3}$ mag airmass$^{-1} X$, and
$\hat m_{6,7}$ equals $m_{6,7}$ minus the median of its values 
after contact 4 on the transit night.
Column 5 is the instrumental B-band magnitude of HD 80607 with arbitrary zero point.
Columns 6 and 7 are the (x, y) positions of HD 80606 in pixels.
Columns 8 and 9 are the (x, y) positions of HD 80607 in pixels.
Five lines demonstrate the format of the table; the complete table of 5708 lines
is available online.}
\end{deluxetable*}

\begin{deluxetable*}{ccccccccccccccc}
\tablewidth{0pt}
\tabletypesize{\small}
\tablecaption{Transit Models \label{table:models} }
\tablehead{\colhead{Model} & \colhead{\RpRs} & \colhead{e} & \colhead{i} & \colhead{$z_{\rm min}$} & \colhead{$t_1$} & \colhead{$t_2$} & \colhead{$<t>$} & \colhead{$t_3$} & \colhead{$t_4$} & \colhead{$F_B(z_{\rm min})$} & \colhead{$F_R(z_{\rm min})$} & \colhead{$\chi^2_\nu$} \\
&       &         &  \colhead{[deg]} &  \colhead{[$R_*$]} &         &         &         &         &         &        &         &
}
\startdata
A &  0.11 &  0.9337 &  89.26 &  0.847 &   0.099 &   0.243 &   0.332 &   0.421 &   0.565 &  0.9898 &  0.9891 &  1.00 \\ 
B &  0.11 &  0.9331 &  89.21 &  0.909 &   0.184 &   --    &   0.391 &   --    &   0.598 &  \underbar{\it 0.9918} &  0.9908 &  \underbar{\it 1.82} \\ 
C &  0.11 &  0.9346 &  89.42 &  0.658 &  \underbar{\it -0.045} &   0.050 &   0.245 &   0.439 &   0.534 &  \underbar{\it 0.9868} &  0.9870 &  \underbar{\it 1.69} \\ 
D &  0.09 &  0.9345 &  89.50 &  0.568 &  \underbar{\it -0.046} &   0.025 &   0.256 &   0.487 &   0.558 &  0.9905 &  0.9909 &  1.36 \\ 
E &  0.15 &  0.9334 &  89.17 &  0.952 &   0.150 &   --    &   0.360 &   --    &   0.570 &  0.9895 &  \underbar{\it 0.9881} &  1.00 \\ 
\enddata
\tablenotetext{}{Times of contacts and their mean $<t>$ in columns 6-10 equal HJD - 2454876.
Columns 11 and 12 are the minima of the normalized fluxes in the B and R bands respectively. 
Column 13 is the B-band reduced chi-squared. Underlined italics identify constraining values
discussed in the text.}
\end{deluxetable*}

\begin{figure}
\plotone{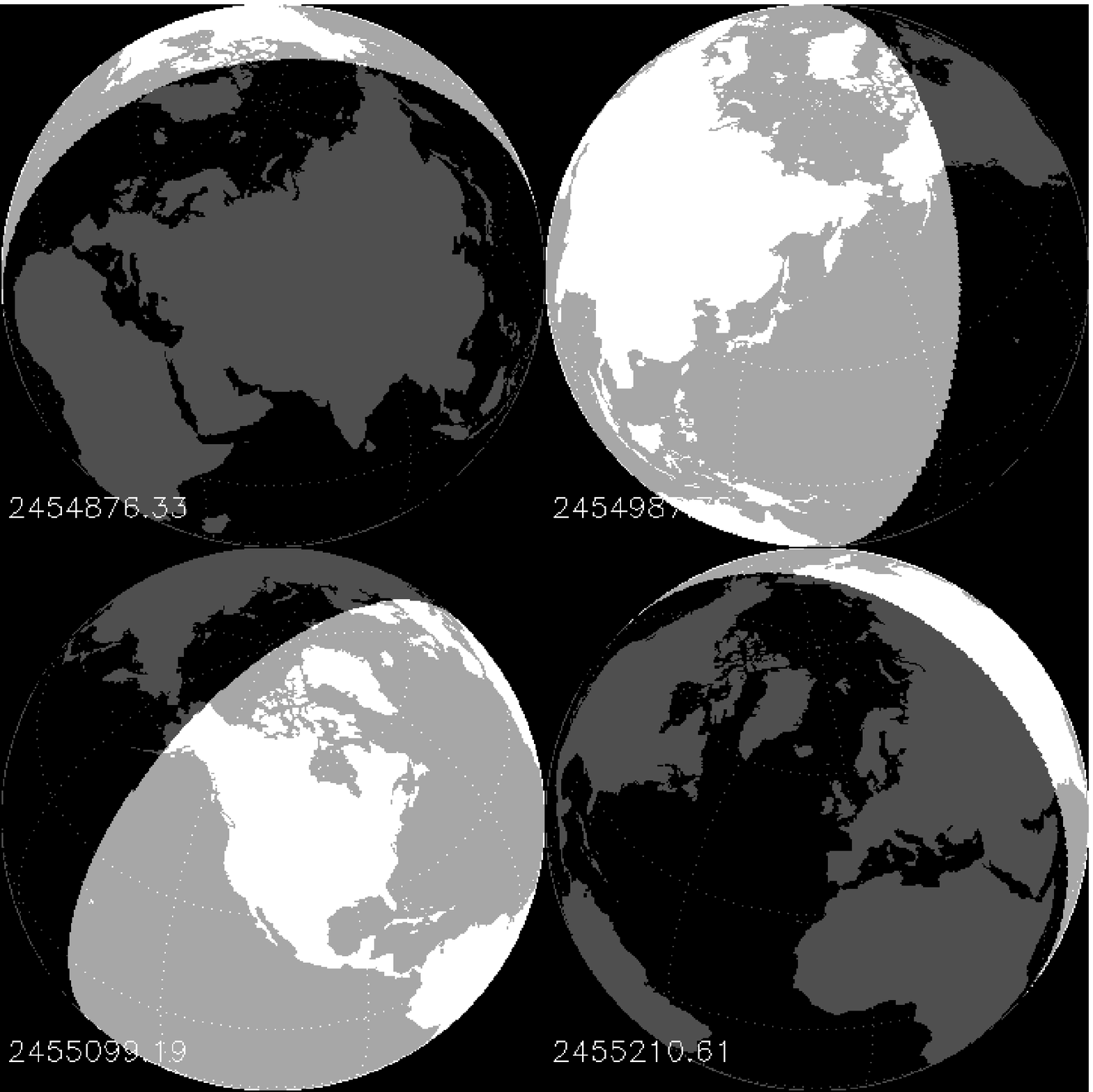}
\caption{Views of Earth from HD 80606b at mid-transit. The predicted HJD
is labeled.
Areas in which the Sun is above the local horizon are brighter.
The next two opportunities to observe a transit of HD 80606b from the surface of the
Earth (upper right and lower left) are not as favorable in terms of daylight
and local airmass as the one
reported here (upper left) or the one 334 days later (lower right).
\label{fig:views}}
\end{figure}

In principle, comparison of our B-band data with data obtained through a redder filter 
could help eliminate degeneracies inherent to
single-color transit light curves (see e.g. the analysis of TrES-3 by
O'Donovan et al. (2007)). 
Extant R-band light curves (Fossey et al. 2009; Moutou et al. 2009) could serve this purpose,
although it appears that for an unambiguous comparison, additional observations will be required.
In practice, very careful calibration is required for such a comparison,
because transit light curves are nearly achromatic. 
For mis-calibration, the possibilities are myriad and the effects significant.
In the small limb darkening approximation, the depth $1-F(z_{\rm min})$ of a non-grazing transit 
is $\sim$ (\RpRs)$^2$. A fractional calibration error in $F(z_{\rm min})$ of $\epsilon$
results in an error in \RpRs\ of $\sim -0.5 \epsilon$ (\RpRs)$^{-2}$.
For the case of HD 80606b, with \RpRs$ \approx 0.1$,
if $\epsilon \approx +0.002$, then the uncertainty in \RpRs\ is -10\%. 
A potential contributor to $\epsilon$ is the uncertainty in the
out-of-transit level, which often is measured at larger airmass
over a shorter period of time than the in-transit observations. In
this case, after fourth contact we observed 820 points, each with
0.63\% uncertainty, so the uncertainty of their average must be at
least $0.22\times10^{-3}$, i.e. a 1.1\% uncertainty in \RpRs.
Often nonlinear detector response may affect directly the measured
depth of a transit, but in this case, we limited the maximum counts
to a small fraction of the full well of a pixel. Thus, we expect
nonlinearity to be insignificant, an expectation supported by the
``control'' time series at HJD 2454881.58, at which time clouds
increased the instrumental magnitude by $\sim 2$ mag,
but the differential magnitude of the two stars, nominally -0.15 mag,
was unaffected.

The wiggles at the beginning (also bottom) of the
light curve may be either due to poorer calibration at large airmass, or
due to non-uniformities (spots) on the surface of the star, HD 80606.
Spots can affect details of light curves and their interpretation. Spots also
could affect radial velocities obtained during and adjacent to the transit.
The Rossiter effect for HD 80606b may indicate significant inclination between
the orbital plane and the equator of the star (Moutou et al. 2009).
If so, then careful observations of transits could provide means of
measuring the latitudinal distribution of spots on HD 80606 and
of monitoring variations in that distribution over the years.

\section{Summary}

We report a detection of a transit of HD 80606b observed in B band. 
The transit depth and egress duration are consistent with expectations.
We refine the orbital eccentricity,
$e = 0.9337^{+0.0009}_{-0.0006}$, and the orbital inclination, $i = 89.26^{+0.24}_{-0.09}$ degrees.
The planetary radius in units of the stellar radius is \RpRs\ = $0.11^{+0.04}_{-0.02}$.

Although less complete than we would have liked, 
the suite of observations of HD 80606b's Valentine's transit, i.e. two-color
photometry and high-resolution spectroscopy,
is much more thorough than typical for a planet on its first witnessed transit.
Observations of a complete transit, from before ingress to after egress, would improve significantly
the measurement of the planetary radius.
Figure \ref{fig:views} illustrates the Earth as it would appear
from HD 80606 at the times of mid-transit for the transit reported
here and the next three transits.  Figure \ref{fig:views} illustrates
a benefit of being above the Earth's bright atmosphere for observing
extrasolar planets with orbital periods much larger than a month.
Many planets discovered by {\it Kepler} will have similarly infrequent opportunities
for Earth-bound observation.

\acknowledgments

P. R. M. is funded primarily by NASA Origins of Solar Systems grant NNG06GG92G.

\end{document}